\documentclass{PoS}

\usepackage{epsfig}

\title{Ghost-gluon and ghost-quark bound states and their role in BRST quartets}

\ShortTitle{Ghost-gluon and ghost-quark bound states and their role in BRST quartets}

\author{\speaker{Natalia Alkofer}
\\
Institut f\"ur Physik,
Karl-Franzens-Universit\"at,
Universit\"atsplatz 5,
A-8010 Graz, Austria\\
       E-mail: \email{natalia.alkofer@edu.uni-graz.at}}

\author{Reinhard Alkofer\\
Institut f\"ur Physik,
Karl-Franzens-Universit\"at,
Universit\"atsplatz 5,
A-8010 Graz, Austria\\
E-mail: \email{reinhard.alkofer@uni-graz.at}}

\abstract{A non-perturbative version of the BRST quartet mechanism in 
infrared Landau gauge QCD is proposed for transverse gluons and quarks. 
Based on the positivity violation for transverse gluons the content of 
the respective non-perturbative BRST quartet is derived. To identify 
the gluon's BRST-daughter and second parent state, a truncated 
Bethe-Salpeter equation for the gluon-(anti-)ghost bound state is 
investigated. 
We comment shortly on several equivalent forms of this equation. 
Repeating the same construction for quarks leads to a truncated 
Bethe-Salpeter equation for a fundamentally charged quark-(anti-)ghost 
bound state. It turns out that a cardinal input to this equation is 
given by the fully dressed quark-gluon vertex, and that it is indispensable 
to dress the quark-gluon vertex in this equation in order to obtain a 
consistent truncation.  }

\FullConference{International Workshop on QCD Green's Functions, Confinement and Phenomenology\\
 5-9 September 2011\\
 Trento, Italy}

\begin{document}

% ------------------------------------------------------------------------------------------------------

\section{Introduction}

The topic of this workshop is  QCD Green's functions as well as their relation
to confinement and  the corresponding phenomenology.  Amongst these Green's
functions  the gluon propagator is one of (if not the) most investigated one.
It has been calculated with various methods in various gauges. In Landau  gauge,
where all gluons contributing to physical states are strictly  transverse and
therefore in the massless representation of the  Poincar{\'e} group, it is
described by one renormalisation function  $Z(k^2)$. Already in the late
seventies it has been noted that there is  a conflict between
antiscreening\footnote{leading to the celebrated  asymptotic freedom property}
by and positivity of transverse gluons  \cite{Oehme:1980ai}.  Nowadays, thirty
years later, there is no doubt any more that  the gluon propagator of Landau
gauge QCD violates positivity,  see {\it e.g.\/} Ref.\ \cite{Bowman:2007du}  and
references therein. Due to this property we can conclude that the
one-gluon-state (with a transverse gluon)  belongs to the states of negative
norm in the state space of Landau gauge QCD. This is in contrast to the
characterization of the indefinite-metric state space of gauge theories based on
perturbation theory (or, expressed more formally, on the limit of vanishing
gauge coupling): then only the longitudinal and the time-like component of the
gauge boson (as well as ghosts and antighosts) are related to the negative and
zero norm states \cite{Peskin:1995ev,Weinberg:1996kr}. One possible route is 
then to proceed by
establishing a Hilbert space of Becchi-Rouet-Stora--Tyutin (BRST) singlets 
via the construction of a BRST cohomology, see {\it e.g.\/}
\cite{Nakanishi:1990qm,Weinberg:1996kr}.

It is evident that in a full solution to Landau gauge QCD in this way all colour
charged particles will be removed from the physical spectrum because the BRST
singlets of gauge-fixed theory are gauge-invariant states. Note that this
construction does not tell us how the removal of non-invariant states occurs
dynamically, it is even up to today unclear how the details of a
non-perturbative BRST cohomology might work. Within the context of Landau gauge
QCD we formulated some first steps into this direction
\cite{Alkofer:2011pe}\footnote{For a presentation of this topic in  a more
introductory style we refer to \cite{Alkofer:2011uh}.}.  The essential point to
be noted is the way the non-perturbative character of the BRST multiplets
containing the transverse gluon, resp., the quark is realized:
These quartets are besides the transverse gluon, resp., quark build of bound
states. It is here where the full difficulty on non-Abelian gauge theory (and
therefore QCD) strikes back: The solution of relativistic bound state equations
in a symmetry-preserving way is a notoriously complicated problem. 

As will be detailed below the above remarks on positivity violation  in the
gluon propagator especially imply that the quantum state of one transverse gluon
can be identified with a parent state in a BRST quartet. As we stated above the
other members, however, have to be non-perturbative, {\it i.e.\/} bound states.
We will present the identification of possible members of this quartet and describe
a strategy to provide evidence for their role in the formalism of covariantly
gauge-fixed  Yang-Mills (YM) theory. 
If this approach turns out to be successful it will provide further details  on
the kinematical aspects of gluon confinement in Landau gauge. 

In the quark sector of Landau gauge QCD the situation is even much less clear.
To the best of our knowledge there is no convincing evidence on the issue of
positivity violation  for the quark propagator in either direction,
positivity-violating or positivity-respecting.  Following a similar strategy
as for the gluons  may allow a clarification whether quarks are also positivity
violating. Therefore, we will present also the case of constructing a 
(hypothetical) quark BRST quartet.

% ------------------------------------------------------------------------------------------------------

\section{Non-perturbative BRST quartets in Landau gauge QCD}

A basic understanding of the BRST quartet cancellation mechanism can be gained
from analysing the so-called elementary quartet
\cite{Nakanishi:1990qm,Weinberg:1996kr}.  Referring to the gauge fields
generically as gluons\footnote{Of course, this procedure applies to all YM
theories in Faddeev-Popov quantization.} the elementary quartet consists of
longitudinal and time-like gluons as well as Faddeev-Popov (FP) 
ghosts and antighosts\footnote{ One
cannot consider directly the longitudinal and time-like gluons but  needs to
define linear superpositions of them, the forward, resp., backward polarized
gluons,  see {\it e.g.} Chapter 16 of Ref.~\cite{Peskin:1995ev} for a definition
of these states.}.  
We start with the definition of the BRST transformation $\delta_B$:
\footnote{
The nilpotency of the BRST transformation is explicit in a
representation with Nakanishi-Lautrup field $B^a$.
It becomes on-shell
identical to the  gauge fixing condition, $B^a=({1}/{\xi}) \, \,\partial_\mu
A_\mu^a$, with $\xi$  being  the gauge parameter.}
\begin{equation}
\begin{array}{ll}
\delta_B A^a_\mu \, =\,  \widetilde Z_3 D^{ab}_\mu c^b \, \lambda  \; , 
\quad &  
\delta_B q
\, = \,-  i g t^a \widetilde Z_1 \, c^a \, q \, \lambda \; , \\
\delta_B c^a \, = \, - \, \frac{g}{2} f^{abc} \widetilde Z_1 
 \, c^b c^c \, \, \lambda \; ,
\quad  & \delta_B \bar c^a \, = \, B^a
\, \lambda \; , 
\qquad \qquad 
\delta_B B^a \, = \, 0,\end{array}  
\label{BRST}
\end{equation}
where $D^{ab}_\mu$ is the covariant derivative. The Grassmann parameter 
$\lambda$ carries ghost number
$N_{\mbox{\tiny FP}} = -1$. $\widetilde Z_1$ and $\widetilde Z_3$ are the
ghost-gluon-vertex and the ghost wave function renormalisation constants.

The BRST charge operator $Q_B$ ({\it e.g.}, derived via Noether's theorem) 
provides then the BRST transform of a field, for example $Q_B A^a_\mu \,
=\,  \widetilde Z_3 D^{ab}_\mu c^b $. It is straightforward to prove that $Q_B$
is nilpotent, $Q_B^2=0$. In addition, as $Q_B$ carries ghost number, its
commutator with  the ghost number operator $Q_c$ is given by 
$\left[i\, Q_c , Q_B \right] = Q_B $. 

In the limit of vanishing coupling $g\to 0$, the elementary BRST quartet can be
read from the BRST transformation in the following way: The forward-polarized
gluon ({\it cf.}\ p.\ 511 of Ref.\ ~\cite{Peskin:1995ev}) is transformed via
$Q_B$ into a ghost. Note that for $g\to 0$ the one-ghost-state is annihilated by
the BRST transformation. The FP-conjugated state of the ghost, the antighost, is
mapped by $Q_B$ to a state of the field $B^a$ whose states are exactly the
backward-polarized gluons. Forward-polarized gluons and antighosts are
BRST non-singlets (and have negative norm), the backward-polarized gluons and
ghosts are in the image of $Q_B$, a property which implies that they have zero
norm. On the other hand, there are non-vanishing off-diagonal matrix elements,
and they are usually for convenience normalized to one. It is this algebraic
structure which guarantees that there are no non-vanishing $S$-matrix elements
between BRST singlets and BRST quartets.

The elementary quartet can serve to explicate the  notation of
Refs.~\cite{Kugo:1979gm,Nakanishi:1990qm}: We call the negative norm state we
start as 1st parent, its BRST transform 1st daughter. The FP-conjugated state
of the latter is then named 2nd parent, the corresponding BRST transform 2nd
daughter.

\begin{figure}[th]
\centerline{\psfig{file=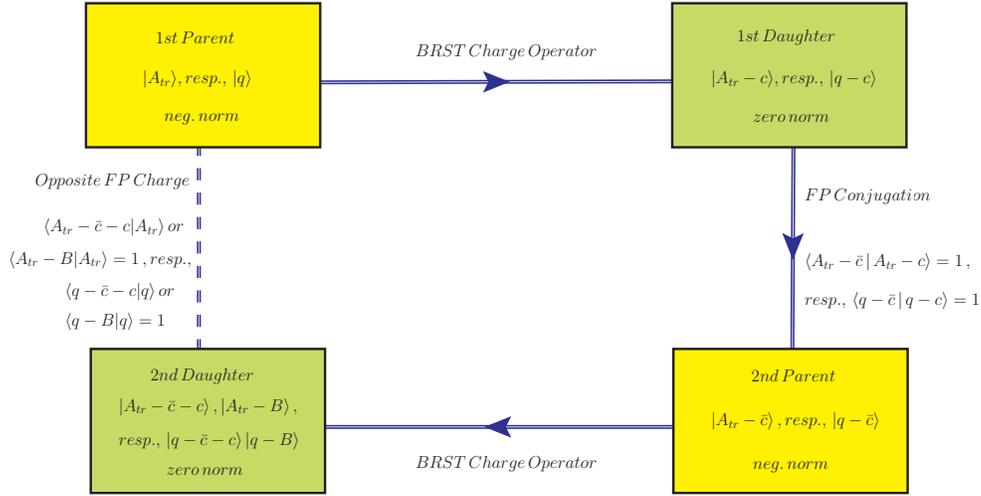,width=130mm}}
\caption{An illustration of the construction of the BRST quartets of the
transverse gluon, resp., quark.
\label{BRSTquartet}}
\end{figure}

The corresponding construction of the BRST quartet starting from a transverse
gluon, resp., a quark is displayed in Fig.~\ref{BRSTquartet}. Starting from the
first transformation in eq.~(\ref{BRST}) and noting already that the term with
the ghost field alone is ``used up'' for the elementary quartet, it is obvious
that the field content of the 1st daughter state is of the form
$  \widetilde Z_3 gf^{abc} A^c_\mu c^b \, .$
Following the construction the 2nd parent is 
$  \widetilde Z_3 gf^{abc} A^c_\mu \bar c^b \, .$
Its BRST transform, the 2nd daughter, is then
$ - \widetilde Z_3^2 gf^{abc} \bar c^b D_\mu^{cd} c^d + \widetilde Z_3 gf^{abc} 
B^b A_\mu^c$. The field content for the quark BRST quartet accordingly reads:

\smallskip 

\centerline{
$q, \quad - \widetilde Z_1 ig t^ac^a q, \quad - \widetilde Z_1 ig t^a \bar 
c^a  q,
\quad  - \widetilde Z_1 ig t^aB^aq - \widetilde Z_1^2 g^2 \frac i 2 f^{abc} t^c
\bar c^ac^b q .
$}

\smallskip

For every ``one-transverse-gluon'' state, resp., ``one-quark'' state  there
should occur exactly one degenerate daughter state. This necessitates the
existence of a ghost-gluon bound state in the adjoint and of a quark-ghost bound
state in the fundamental  representation \cite{Alkofer:2011pe,Alkofer:2011uh}.
The FP-charge reversed 2nd parent state is an antighost-gluon, resp.,
antighost-quark bound state. Landau gauge provides now  an advantage: 
In the limit $\xi\to0$ the formalism becomes
ghost-antighost-symmetric, and thus a ghost-gluon (-quark)
bound state implies a degenerate antighost-gluon (-quark) bound state with same
quantum numbers. The BRST transformation leaves then three, resp., two 
possibilities for the 2nd daughter. In the gluon case they are 
ghost-antighost, ghost-antighost-gluon, or a bound
state of two differently polarized gluons, whereas in the quark case it is a 
ghost-antighost-quark, or a bound state of the backward-polarized
gluon with the quark.

At this point a remark with respect to a fundamental difference between
transverse gluons and quarks is in order. As stated already above the transverse
gluons, as vector states with two polarizations, are in the massless
representation of the Poincar{\'e} group.\footnote{Please note that this does
not exclude the kind of screening mass determined by a non-vanishing infrared
value of the gluon propagator. Such a ``gluon screening mass'' has been 
suggested already thirty years ago in Ref.~\cite{Cornwall:1981zr}
and has been found more recently, {\it e.g.}, in
Refs.~\cite{Aguilar:2008xm,Boucaud:2008ky,Fischer:2008uz}. }
Correspondingly all the other partners of the quartet will appear with two
polarizations and are thus also  in the massless
representation of the Poincar{\'e} group.
Quarks, on the other hand, will be described by massive Dirac fermions. For 
light quarks mostly dynamical chiral symmetry breaking and for heavy
quarks mostly explicit chiral symmetry breaking determines the corresponding
effective masses. In case the quark BRST quartet exists its members will be all,
in the sense of the representation of the Poincar{\'e} group, massive Dirac 
fermions. However, contrary to the situation for the transverse gluons the
question of positivity violation of quarks is unresolved. Although the 
 infrared behaviour of the quark propagator comes out almost identical 
in all the studies done so far its analytic structure remains elusive.
In functional approaches this can be traced back to the fact that 
the analytic structure of the quark propagator is highly
sensitive to details in the quark-gluon vertex,  see, {\it e.g.\/},
Ref.~\cite{Alkofer:2003jj,Aguilar:2010cn}. The quark-gluon vertex is, on the
other hand, also very strongly influenced by dynamical and/or
explicit chiral symmetry breaking
\cite{Skullerud:2003qu,Alkofer:2006gz,Alkofer:2008tt}. 
Already these remarks make evident that the mass generation for
quarks related to chiral symmetry breaking depends strongly on 
details of the dynamics. What kind of mechanism then guarantees that the 
corresponding BRST bound
states are degenerate with the quark states is completely unknown. 
In the following we will have a closer look to the bound state equations for the 
non-perturbative BRST quartet states in the hope that their structure provides
an insight into possible answers to these questions.

% ------------------------------------------------------------------------------------------------------

\section{Ghost-gluon bound state equation}
 
Our aim is to derive a consistently truncated bound state equation for the 1st
daughter of the BRST quartet of the transverse gluon. As stated above this gives
us automatically the bound state equation for the 2nd parent. Determining the
necessary input in terms of primitively divergent YM Green's functions is
certainly a demanding task. However, given the knowledge acquired for the gluon
and ghost propagators as well as for the gluon-ghost and three-gluon vertex
functions over the last decade, a reasonable precise modelling of this input is
available. Therefore the real challenge is to find a trustable truncation.

To this end we exploit a peculiarity of the solutions of Dyson-Schwinger and
Functional Renormalisation Group equations: There is one
unique scaling solution with power laws for the Green's functions 
and an one-parameter family of solutions,
the so-called decoupling solutions. The latter are infrared trivial
solutions which possess as an endpoint exactly the scaling solution
characterized by infrared power laws.  Numerical
solutions of the decoupling type (there called ``massive solution'')
have been published in \cite{Aguilar:2008xm,Boucaud:2008ky} and
references therein. A recent detailed description and comparison of
these two types of solutions has been given in Ref.\
\cite{Fischer:2008uz}, see also Refs.
~\cite{Alkofer:2008jy,Huber:2009wh,Szczepaniak:2001rg,Epple:2007ut,Fischer:2009tn}.
This leads to the idea to order the diagrams according to the infrared behaviour
seen in the scaling solution. If the conjecture of Ref.~\cite{Maas:2009se},
namely that the occurrence of different types of solutions is a gauge artefact, 
is correct, it is
sufficient that only one non-perturbative completion of Landau gauge with
scaling solution exists to make the above described truncation 
well-founded. Second, one may argue based on continuity that diagrams which are
leading in the scaling solution will be numerically dominant in the decoupling
solution.

Investigations of the scaling solution have been performed for the gluon and
ghost propagators 
\cite{vonSmekal:1997is,Watson:2001yv,Zwanziger:2001kw,Lerche:2002ep,Fischer:2002hna}
and for Yang-Mills vertex functions
\cite{Alkofer:2004it,Huber:2007kc,Alkofer:2010tq,Alkofer:2011va}. In this
solution every one-particle irreducible Green's functions behaves like a
power-law in the deep infrared. Therefore one can attribute an infrared exponent
to every diagram appearing in a Dyson-Schwinger or Functional Renormalisation
Group equation.  With this in mind the ghost-gluon bound state is looked for in
the ghost-gluon scattering kernel. To this end we want to truncate this quantity
to the infrared leading term.  We use  DoDSE
\cite{Alkofer:2008nt,Huber:2010ne,Huber:2011qr}  to derive the diagrammatic
expressions for the Dyson-Schwinger equation of this four-point function. A
diagram-by-diagram infrared power counting is performed by attributing anomalous
infrared exponents to each of them. A consistency
check is provided by the fact that the a priori known infrared exponent of the
ghost-gluon scattering kernel is refound.

\begin{figure}[th]
\centerline{\psfig{file=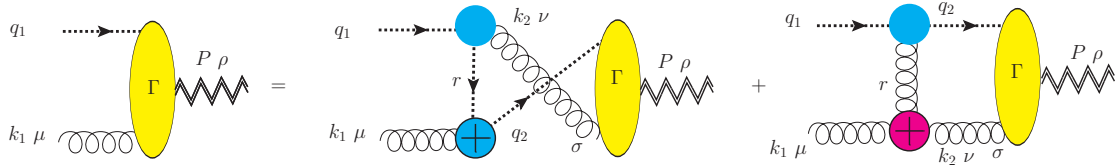,width=150mm}}
\caption{Graphical representation of the ghost-gluon Bethe-Salpeter equation.
Crosses denote dressed vertices.
\label{GhGl}}
\end{figure}

To select the diagrams to be kept in the bound state equation we require:
They should contain
the one-particle irreducible ghost-ghost-gluon-gluon four-point function and no
$n\ge 5$-point function, they should be infrared leading, and
the interaction shall take place in the ghost-gluon channel.
This leaves from the many  diagrams in the equation for the ghost-gluon
scattering kernel only two:
One with two ghost and one gluon propagators on
internal lines. This is effectively a ghost exchange.
And another one with two gluon and one ghost propagators on
internal lines. This is a gluon exchange. This diagram is also infrared
leading because in the scaling solution the fully dressed
three-gluon vertex is infrared divergent.

Assuming the existence of a bound state as well as employing the usual
decomposition of the (ghost-ghost-gluon-gluon) four-point function into
Bethe-Salpeter amplitudes and performing the expansion around the pole (see {\it
e.g.\/} Sect.~6.1 of Ref.~\cite{Alkofer:2000wg}) one arrives at the
Bethe-Salpeter equation depicted in Fig.~\ref{GhGl}. Using the propagator
parameterizations of {\it e.g.\/} Ref.\ \cite{Alkofer:2003jj}, the ghost-gluon
vertex of Ref.\ \cite{Schleifenbaum:2004id}, and the three-gluon vertex of Ref.\
\cite{Alkofer:2008dt}, one can derive a self-consistent equation for the
corresponding Bethe-Salpeter amplitude containing otherwise only known
quantities.

As stated above, and as explicitly demonstrated in ref.~\cite{Alkofer:2011pe},
the Bethe-Salpeter amplitude for this bound state in the adjoint colour
representation is transverse, {\it i.e.}, this bound state is as expected in the
massless representation of the Poincar{\'e} group. Restricting oneself to the
(presumably dominant)  ghost exchange and going to the limit of soft momenta,
in  ref.~\cite{Alkofer:2011pe} the following equation for the ghost-gluon
Bethe-Salpeter amplitude (being described by one function $F$  as only the
transverse component exists) is obtained:
\begin{equation}
F(k_1^2) =  \widetilde Z_3^2 g^2 N_c^2 
\int \frac {d^4k_2}{(2\pi)^4} \frac{G((k_1+k_2)^2)}{(k_1+k_2)^2}
\frac{G(k_2^2)}{k_2^2} \frac{Z(k_2^2)}{k_2^2}
 \frac 1 3 (k_1\cdot k_2) 
\left( 1 - \frac {(k_1\cdot k_2)^2}{k_1^2 k_2^2}\right) F(k_2^2), 
\label{GhGlBSforF}
\end{equation}
where $Z$ and $G$ are the gluon and ghost renormalisation function, 
respectively. Note that the product $G^2Z$ appearing in the integrand 
is a renormalisation group invariant \cite{Alkofer:2002ne}. 
Using an alternative projection on Lorentz indices one can derive an equivalent,
somewhat simpler, equation:
\begin{eqnarray}
F(k_1^2) = -\frac 1 3 k_1^2  Z_3^2 g^2 N_c^2 
\int \frac {d^4k_2}{(2\pi)^4} \frac{G((k_1+k_2)^2)}{(k_1+k_2)^2}
\frac{G(k_2^2)}{k_2^2} \frac{Z(k_2^2)}{k_2^2}    
\left( 1 - \frac {(k_1\cdot k_2)^2}{k_1^2 k_2^2}\right) F(k_2^2).  
\label{GhGlBSforF2}
\end{eqnarray}
Attempts to numerically solve these equations by iteration with the input as 
described above have generated curves which deviate from each other in the 
infrared. Also for both curves some unexpected features in deep infrared 
occur. Therefore we decided to change the numerical method, work in this
direction is in progress.

% ------------------------------------------------------------------------------------------------------

\section{Ghost-quark bound state equation}

The quark propagator is due to dynamical or explicit chiral symmetry breaking
infrared finite. In the scaling solution, the twelve possible Dirac tensor 
structures of the quark-gluon vertex are then all infrared divergent. This 
leads to an
$1/k^4$ behaviour of the kernel in the four-quark function, $k$ being the
momentum exchange. This is indicative of a linearly rising  potential between
static quarks, and thus quark confinement.

\begin{figure}[th]
\centerline{\psfig{file=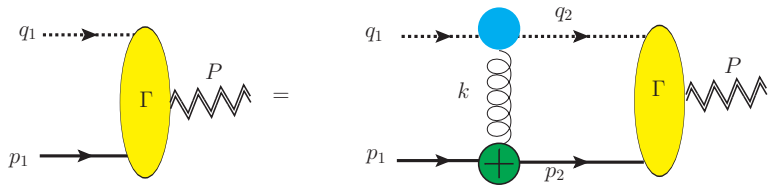,width=105mm}}
\caption{Graphical representation of the ghost-quark Bethe-Salpeter equation.
Here the quark-gluon vertex is fully dressed.
\label{GhQu}}
\end{figure}

Using the same truncation  requirements and the same derivation of the
Bethe-Salpeter equation as in the previous subsection one arrives at the
equation depicted in Fig.~\ref{GhQu} \cite{Alkofer:2011pe}. This equation is in
full agreement with the infrared analysis of the scaling solution.
In its kernel the anomalous
infrared exponents cancel if and only if the fully dressed quark-gluon vertex is
taken into account. Again this is in agreement with the expectations arising from
previous studies.

% ------------------------------------------------------------------------------------------------------

\section{Outlook}

In this work we reported on a suggestion how to study  the non-perturbative BRST
quartets generated by transverse gluons and quarks quantitatively. Attempts of a
numerical solution of the ghost-gluon Bethe-Salpeter equation uncovered so far
only its sensitivity to the deep infrared. Within this project many open
questions still remain: What are the bound states representing the respective
2nd daughters? Is BRST dynamically broken? Can we answer in this way the
question of  positivity or positivity violation for quarks? And does all this
relate to quark confinement?

% ------------------------------------------------------------------------------------------------------

\section*{Acknowledgments}
We thank the organizers of  {\it QCD--TNT--II}, and  especially Daniele
Binosi and Joannis Papavassiliou,  for all their efforts which made this highly
interesting workshop possible.
The figures here have been drawn with JaxoDraw 
\cite{Binosi:2003yf,Binosi:2008ig}, 
we want to take the opportunity  to thank the authors. 

% ------------------------------------------------------------------------------------------------------

\end{document}